\documentclass[a4paper]{jpconf}
\usepackage[varg]{txfonts}

\usepackage{amsmath}
\usepackage{graphicx}
\usepackage{color}
\usepackage{amsbsy}
\usepackage{amssymb,amstext}
\usepackage[square,sort&compress,comma,numbers]{natbib}
\usepackage{endnotes}

\begin{document}
\title{Orbital migration and Resonance Offset of the Kepler-25 and K2-24 systems}
\renewcommand{\thefootnote}{\star}

\author{C. Charalambous$^1$,
X. S. Ramos$^{1}$\footnote{Present address:
Niels Bohr International Academy, The Niels Bohr Institute, Blegdamsvej 17, DK-2100, 
Copenhagen \O, Denmark}, P. Ben\'{\i}tez-Llambay$^2,$ and C. Beaug\'e$^1$}
\address{$^1$ Instituto de Astronom\'\i a Te\'orica y Experimental, Observatorio Astron\'omico,
Universidad Nacional de C\'ordoba, Laprida 854, X5000BGR, C\'ordoba, Argentina}
\address{$^2$ Niels Bohr International Academy, The Niels Bohr Institute, Blegdamsvej 17, 
DK-2100, Copenhagen \O{}, Denmark}
\ead{charalambous@oac.unc.edu.ar}
\begin{abstract}
Based on the model described in \cite{Ramos2017}, we present an analytical+numerical study of 
the resonance capture under Type-I migration for the Kepler-25 \cite{2014ApJS..210...20M} and 
K2-24 \cite{2016ApJ...818...36P} {\it Kepler} systems, both close to a 2/1 mean-motion 
resonance. We find that, depending on the flare index and the proximity to the central star, the 
average value of the period-ratio between two consecutive planets show a significant deviation 
with respect to the resonant nominal value, up to values well in agreement with the observations.
\end{abstract}
\section{Introduction}
Most of the planets found in {\it Kepler} multi-planetary systems lie outside mean-motion 
resonances (MMRs). This seems incompatible with a formation process strongly affected by planet 
migration and may either indicate that planets formed in-situ (e.g. \cite{2012ApJ...751..158H, 
2013ApJ...770...24P,2014ApJ...780...53C}) or that planetary migration occurred in a turbulent 
environment, where resonance capture is not guaranteed (e.g. 
\cite{2012MNRAS.427L..21R,2013ApJ...778....7B}). However, the {\it Kepler} population also shows 
the existence of a statistically significant number of planetary pairs close to resonances, 
where the orbital period ratios is usually larger than the nominal value. It is generally 
believed that these systems are near-resonant and located outside the libration domain. 

The origin of these near-resonant systems is a dilemma. Planetary migration in a turbulent disk 
\cite{2013MNRAS.434.3018P} can lead to near-resonant configurations, although it is not clear 
whether this mechanism can reproduce the observed near-resonant distribution (e.g. 
\cite{2013MNRAS.435.2256Q}). This mechanism only seems to work for sub-Jovian bodies, and 
therefore, larger than the expected size for most of the {\it Kepler} systems.

In \cite{Ramos2017} we presented an analytical model that allow us to reproduce the general 
trend of the resonance offset, as the disk is assumed significantly flared and with a small 
scale height. This model strongly depends on the planetary masses, values which are not usually 
known for the {\it Kepler} systems, thus we performed Monte Carlo statistical analysis 
with no direct application to a given planetary system. Even so, the method has proved promising 
and able of reproducing the offset distribution around both the 2/1 and 3/2 MMRs, as well as 
predicting an increase in this value for planets closer to the central star.
\begin{figure}
\centering
\includegraphics[width=0.45\textwidth]{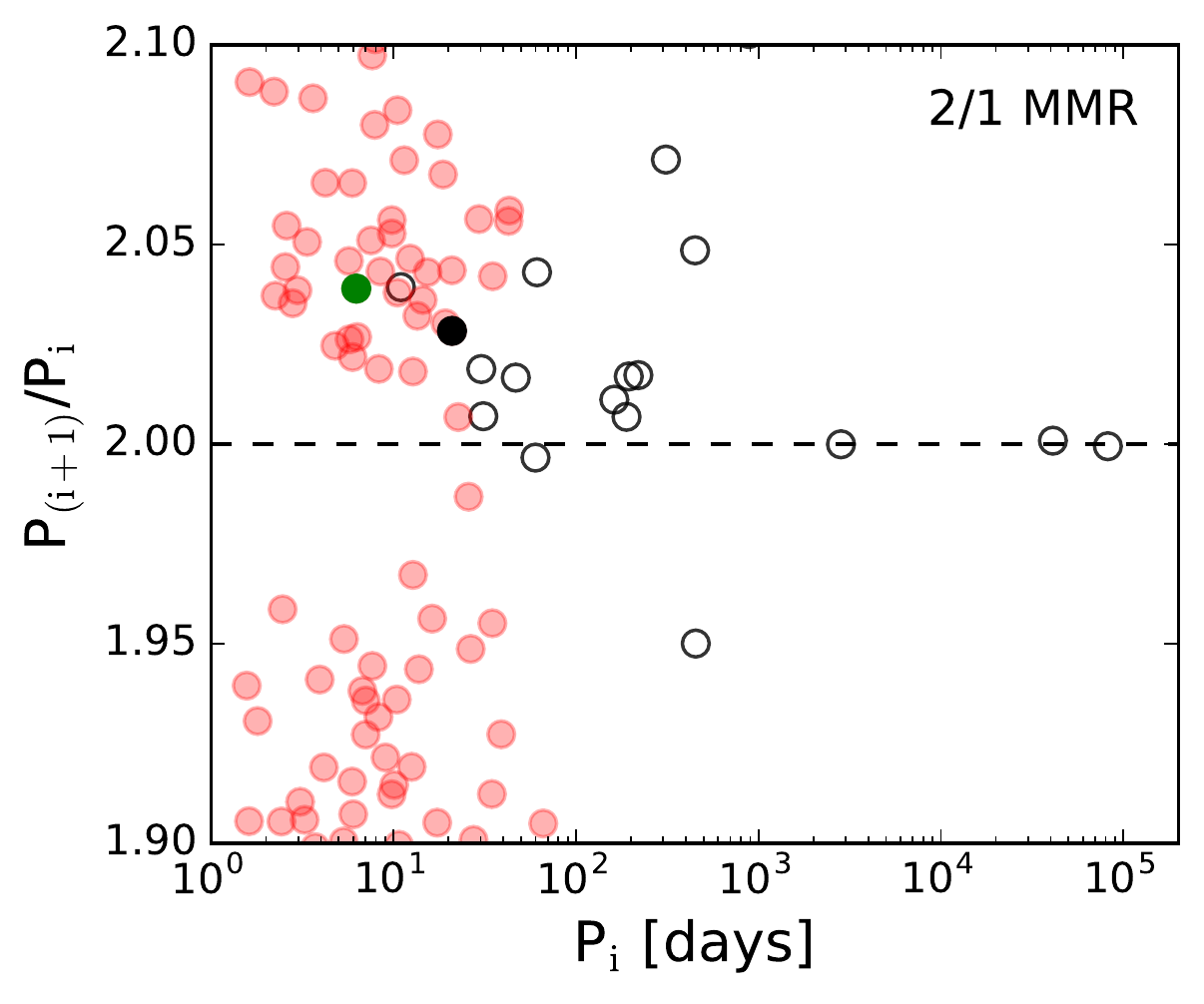}
\includegraphics[width=0.45\textwidth]{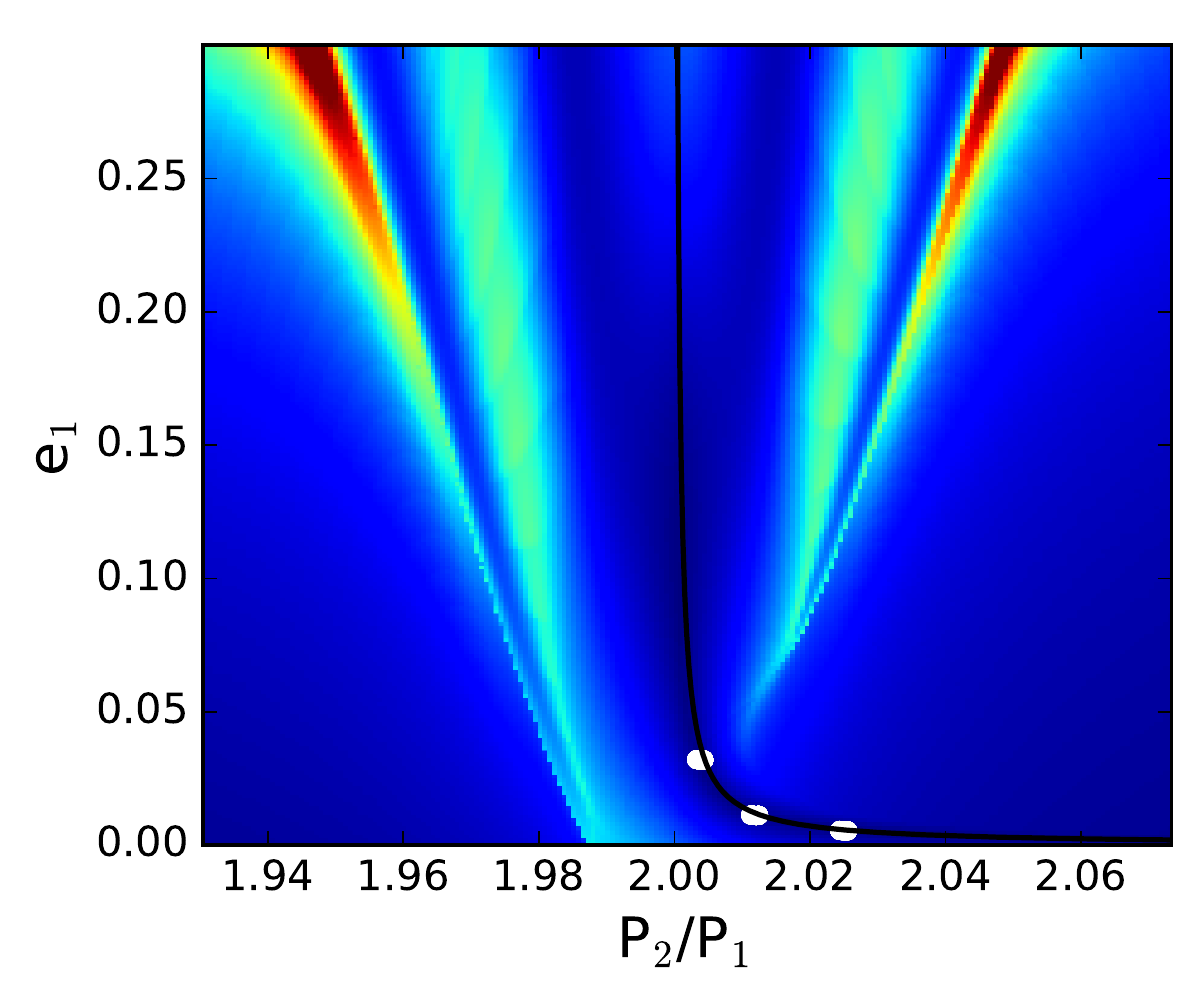}
\caption{{\bf Left:} Distribution of orbital period ratios, in the vicinity of the 2/1 
resonance, as a function of the orbital period of the inner planet. Red circles identify planets 
detected by transits or TTV, while those discovered by other methods are depicted with open 
circles. The green circle indicates the location of Kepler-25 while the black one corresponds to 
K2-24. Data was obtained from {\tt exoplanet.eu}. {\bf Right:} Dynamical map of $max(\Delta e)$ 
for two-planet systems with $m_1 = 0.05 m_{\rm Jup}$ and $m_2 = 0.10 m_{\rm Jup}$ orbiting a 
central star of mass $m_0 = 1 M_{\odot}$, in the vicinity of the 2/1 MMR. The orbit of the outer 
planet was initially circular with $a_2=1$ AU, and all the angular variables where chosen equal 
to zero. The black continuous line marks the location of the zero-amplitude ACR-type librational 
solutions, estimated from the simple analytical model (eq. (\ref{eq15})). White dots are the 
result of three N-body simulations of resonance trapping.}
\label{fig1}
\end{figure}

From all the {\it Kepler} systems near the 2/1 MMR, just two cases have fairly credible 
estimations for the masses: Kepler-25 and K2-24, indicated as green and black circles in the 
left hand frame of Figure \ref{fig1}. Red circles correspond to systems detected by transits or 
Transit Time Variations (TTV). Bodies discovered by any other method are identified by open 
circles. This distribution shows increasing $\Delta_{2/1}$ for planets closer to the star, 
indicating a possible smooth trend which, if confirmed, would indicate that the distribution 
found in different populations belong to the same functional form, and just differing from the 
distance to the star. It also shows little correlation with either the detection method or the 
stellar/planetary masses. Here, we apply the model of \cite{Ramos2017} to these two systems and 
attempt to constraint the properties of the protoplanetary disk that are consistent with their 
observed location and deviation from the exact resonance.

Right frame of Figure \ref{fig1} shows a dynamical map for the 2/1 commensurability. We 
integrated series of two-planet systems with initial conditions in a grid defined in the 
$(P_2/P_1,e_1)$ plane and specifically chose $m_2/m_1 > 1$ to guarantee symmetric fixed points 
for the resonant angles (e.g. \cite{2006MNRAS.365.1160B, 2008MNRAS.391..215M}). The color code 
corresponds to the maximum value of $|e_1(t) - e_1(t=0)|$ (denoted as $max(\Delta e)$) attained 
a during $10^3$ years integration time. Darked (lighter) tones are associated to small (large) 
variations in the eccentricity of the inner planet. Although this indicator does not measure 
chaotic motion, it is an important tool to probe the structure of resonances and identify the 
locus of stationary solutions (so-called ACR solutions, see \cite{2003ApJ...593.1124B, 
2006MNRAS.365.1160B}). It also helps to identify the separatrix delimiting the librational from 
the circulation domains (e.g. \cite{2015CeMDA.123..453R}). The black line shows the approximate 
location of the family of zero-amplitude ACR solutions characterized by the simultaneous 
libration of the both resonant angles.

\section{The Ramos et al. Model}

In the following we will assume two planets of masses $m_1$ and $m_2$ orbiting a star $m_0$, 
orbital periods $P_1 < P_2$ and in the vicinity of a first-order $(p+1)/p$ MMR. We define the 
{\it resonance offset} $\Delta_{(p+1)/p}$  as 
\begin{equation}
\Delta_{(p+1)/p} = \frac{P_2}{P_1} - \frac{(p+1)}{p},
\end{equation}
whose value indicates the distance from the exact resonance. 

Different values of $\Delta_{(p+1)/p}$ are attained in different parts of the disk. In order to 
study this, we use a resonant Hamiltonian neglecting secular perturbations to estimate the 
resonance offset as function of $e_1$, as well as a relation between the eccentricities of both 
planets: 
\begin{equation}
\Delta_{(p+1)/p} = C_1(\alpha) \; \frac{m_2}{m_0} \frac{1}{e_1} 
\hspace*{0.5cm} ; \hspace*{0.5cm} e_2 = C_2(\alpha) \; \frac{m_1}{m_2} e_1,
\label{eq3}
\end{equation}
(see \cite{1988AJ.....96..400F,2004ApJ...611..517L}), where the coefficients $C_i$ depends 
solely on $\alpha = a_1/a_2$ ($C_1 \simeq 1.5$ and $C_2 \simeq 0.29$). For a given resonance, 
very low $e_i$ are necessary to obtain a significant deviation from the exact resonance. 
However, since $e_i$ does not attain zero for the ACR solution, the singularity at $e_1=0$ is 
never reached.

If the disk-driven planetary migration is sufficiently slow and smooth, we expect the orbital evolution to follow the pericentric branch into the librational domain and exhibit low-amplitude oscillations of the resonant angles. In such an ideal scenario, the final eccentricities and resonant offset $\Delta_{(p+1)/p}$ will depend on the relative strength between the eccentricity damping and orbital migration timescales (\cite{2002ApJ...567..596L,2006MNRAS.365.1160B}) 
$\tau_{e_i}$ and $\tau_{a_i}$, respectively. Thus, the final outcome of a resonance trapping will depend on the ratios ${\cal K}_i = \tau_{a_i}/\tau_{e_i}$, which we denote as the {\it K-factors}. 

We performed 3 N-body simulations including an ad-hoc external acceleration (e.g. \cite{2008A&A...482..677C}), set the values of $\tau_{a_i}$ at certain prefixed amounts, varied ${\cal K}_i$ and analyzed its effects on the resonance offset (white dots in the right frame of Figure \ref{fig1}). We chose planetary masses $m_1 = 0.05 m_{\rm Jup}$ and $m_2 = 0.10 m_{\rm Jup}$, and $1M_\odot$ for the central star. The outer planet was initially at $a_2=1$ AU, in circular orbit and all angles equal to zero. The initial conditions were integrated for $10^5$ yrs. All values agree with the ACR loci given by expression (\ref{eq3}) deduced from the analytical resonance model. Although these simulations indicate that is possible to obtain large values for the offset, they only appear attainable for K-factors of the order of $10^4$, much higher than predicted by linear models of Type-I disk-planet migration (e.g. \cite{2000MNRAS.315..823P, 2004ApJ...602..388T,2008A&A...482..677C}). In \cite{Ramos2017}, we found that it is possible to overcome this problem assuming a significant flare for the disk (of 
the order of $f \simeq 0.25$) in addition to a relatively small value for the disk aspect ratio ($H_0 \simeq 0.03$). This combination, in addition to moderately low values for the mass ratios of the planets, generates large deviations from exact resonance even with K-factors of the order of $10^2$, well within the classical limits.

We assume a laminar disk with surface density $\Sigma(r) = \Sigma_0 r^{-\alpha}$ and aspect-ratio $H_r(r) = H_0 r^{f}$, where $r$ is the distance to the central star in astronomical units. We will consider $H_0$, $\alpha$ and $f$ as unknown parameters that will be estimated in accordance with the observed dynamical characteristics of the planetary systems.

Following \cite{2002ApJ...565.1257T} and \cite{2004ApJ...602..388T}, orbital migration and 
eccentricity damping timescales are approximated as
\begin{equation}
\tau_{a_i} = Q_a \frac{t_{{\rm wave}_i}}{H_{r_i}^2} 
\hspace*{0.5cm}  ; \hspace*{0.5cm} 
\tau_{e_i} = Q_e \frac{t_{{\rm wave}_i}}{0.780}
\hspace*{0.5cm}  ; \hspace*{0.5cm}   
t_{{\rm wave}_i} = \frac{m_0}{m_i} \frac{m_0}{\Sigma(a_i) a_i^2} \frac{H_{r_i}^4}{\Omega(a_i)}.
\label{eq8}
\end{equation}
In these expressions, $H_{r_i}$ is the disk aspect-ratio in the position of each planet and $\Omega(a_i)$ their orbital frequency. $Q_e$ is a constant introduced by \cite{2006A&A...450..833C} in order to reproduce the eccentricity damping rates from hydro-dynamical simulations, while $Q_a = Q_a(\alpha)$ is a function of the surface density profile. Finally, $t_{{\rm wave}_i}$ is the typical timescale of planetary migration.

Recently, the classical Type-I migration models were revised by \cite{2014AJ....147...32G}, who 
considered the contribution of eccentricity damping to changes in the semimajor axis associated 
to (partial) conservation of the angular momentum. They found that the effective characteristic 
timescale for the orbital evolution should actually be given by $\tau_{{a_{eff}}_i} = 
\left(\tau_{a_i}^{-1} + 2 \beta e_i^2 \tau_{e_i}^{-1}\right)^{-1}$, where $\tau_{a_i}$ and 
$\tau_{e_i}$ maintain the same form as equations (\ref{eq8}) and $\beta$ is a factor that 
quantifies the fraction of the orbital angular momentum preserved during the migration. This 
modified migration timescale changes the K-factor, leading to a new ``effective'' form. This 
revised migration model, together with the analytical resonant Hamiltonian led in 
\cite{Ramos2017} to a relation between the disk properties and resonance offset in the form:
\begin{equation}
\Delta_{(p+1)/p}^2 =
\frac{2}{D} \left( C_1 \frac{m_2}{m_0} \right)^2 
\frac{ \left[(1-D(1+\beta)){\cal K}_2 \left( C_2 \frac{m_1}{m_2}\right)^2 + 
(B+D\beta){\cal K}_1 \left( \frac{\tau_{a_2}}{\tau_{a_1}} \right) \right]}
{ 1 - \left( \frac{\tau_{a_2}}{\tau_{a_1}} \right)} ,
\label{eq15}
\end{equation}
where the parameters $B$ and $D$ depend both on planetary mass ratio and the resonance under 
consideration
\begin{equation}
D = \frac{1}{(p+1)} \left( 1 + \frac{a_1}{a_2} \frac{m_2}{m_1} \right)^{-1} 
\;\; ; \;\; 
B = \frac{m_1n_2a_2}{m_2n_1a_1}+D.
\label{eq5}
\end{equation}

The importance of expression (\ref{eq15}) lies in the fact that it shows that large values of 
the offset may be obtained if the denominator is sufficiently small, independently of the 
K-factors. Since the ratio $\tau_{a_2}/\tau_{a_1}$ depends on the planetary mass ratio (as well 
as on parameters $f$ and $\alpha$), it is possible to obtain values of $\Delta_{(p+1)/p}$ 
consistent with the observed planetary systems, as long as the parameters lie within certain 
values. 

\section{Application to Kepler-25 and K2-24 systems}

\begin{figure}
\centering
\includegraphics[width=0.8\textwidth,clip=true]{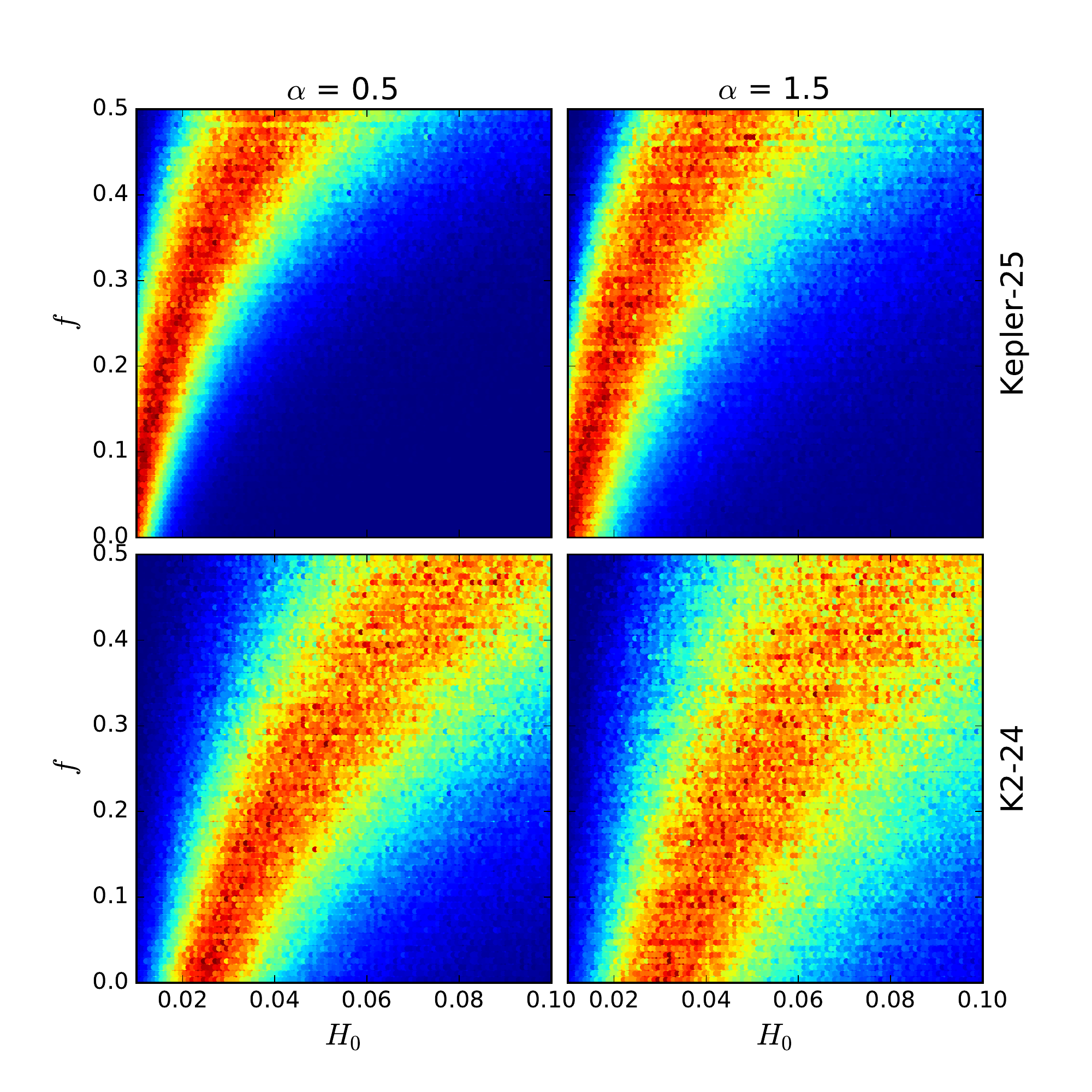}
\caption{Results of Monte Carlo simulations for Kepler-25 (upper frames) and K2-24 (bottom 
panels) for different density slopes values. In the left we show $\alpha=0.5$ and in the right 
$\alpha=1.5$. Red colors indicate more possible pairs $(H_0,f)$ for an observed $\Delta_{2/1}$, 
than the blue values.}
\label{mapas-keplers}
\end{figure}
\begin{table}[h!]
\centering
\caption{Mass measurements and orbital periods for Kepler-25 b,c and K2-24 b,c.}
\begin{tabular}{l r r r r r }
\hline\hline
System &  $m_1$ [$m_\oplus$] &  $m_2$ [$m_\oplus$]  &  $P_1$ [d] & $P_2/P_1$ & $m_0 
[M_\odot]$\\
\hline
Kepler-25  &  $9.6 \pm 4.2$ & $24.6 \pm 5.7$  &  $6.24$  & $2.0390$ & $1.22 \pm 0.06$ \\
K2-24      & $21.0 \pm 5.4$ & $27.0 \pm 7.0$  & $20.89$  & $2.0284$ & $1.12 \pm 0.05$ \\
\hline
\end{tabular}
\label{tab1}
\end{table}

Of 165 planetary pairs lying in the vicinity of the 2/1 resonance and with $P_1 \leq 100$ days, only in 18 cases have the masses of both bodies have been measured or estimated with some accuracy. Of these, only 10 have orbital period ratios in the interval $P_2/P_1 \in [2.0, 2.10]$, and may be thus cataloged as members of the (near)-resonant region. This number continues to decline as we note that 7 planetary pairs have at least one of its members with $m_i > 50 m_\oplus$, more than sufficient to open a gap in the disk and to have migrated following a Type-II scenario. Since our model is based on analytical prescriptions for laminar Type-I migration, these systems are beyond the scope of our work.

Of the three remaining candidates, HD 219134 have recently been questioned. The first reference to this system appears in \cite{2015A&A...584A..72M}, who analyzed a total of 98 nightly averaged RV observations obtained with HARPS-N and found evidence of 4 low-mass planets. The outer member of the (near)-resonant pair was not detected. Later, \cite{2015ApJ...814...12V} analyzed a total of 276 RV Doppler measurements and identified a total of 6 planets in this system. \cite{2016arXiv160205200J} found a substantial periodicity in the RV data due to stellar rotation with a period of 22.8 days, a value practically equal to $P_1/2$. Although the authors do not believe there is sufficient evidence to rule out the existence of the inner planet, the amplitude generated by the planet in the RV signal may be affected by stellar rotation and thus, the mass deduced for $m_1$ could in fact be substantially lower.

This leaves us with two systems, Kepler-25 and K2-24. Table \ref{tab1} gives the masses and orbital periods of both systems, together with the stellar masses and the respective standard deviations. Both systems have nominal mass ratios larger than unity (i.e. $m_2/m_1 > 1$) and are thus candidates for resonant trapping in the 2/1 commensurability. We therefore proceeded to analyze whether the observed value of the resonance offset $\Delta _{2/1}$ could be achieved using our model with the assumption of a laminar flared disk. 

Since the value of $\Delta_{2/1}$ is known, we can invert expression (\ref{eq15}) to obtain explicitly the value of $H_0$ as function of the masses and the disk flare:
\begin{equation}
H_0^2 = \frac{2}{D \, \Delta_{\rm obs}^2} \left( C_1 \frac{m_2}{m_0} \right)^2 
\frac{ \left[(1-D(1+\beta)){\cal K}_2^* \left( C_2 \frac{m_1}{m_2}\right)^2 + 
(B+D\beta){\cal K}_1^* \left( \frac{\tau_{a_2}}{\tau_{a_1}} \right) \right]}
{ 1 - \left( \frac{\tau_{a_2}}{\tau_{a_1}} \right)} ,
\label{eq16}
\end{equation}
where $\Delta_{\rm obs}$ is the observed value of the offset and ${\cal K}_i^* = 0.78 (Q_a/Q_e) a_i^{-2f} $ are $H_0$-normalized expressions for the K-factors of each planet. If the planetary masses are known with even some accuracy, it is then possible to estimate relations between $f$ and $H_0$ leading to the observed values of the offset. Since uncertainties in these values may be significant, we chose a statistical Monte Carlo approach incorporating the errors in the masses into the calculation.

For each system we ran a series of 1000 sets of $(m_1,m_2)$ from a normal distribution with mean and variance as depicted from Table 1. From the values of each run, we then determined the distribution of values of $(H_0,f)$ according to (\ref{eq16}), for fixed values of $\alpha$. Results are shown in color scales in Figure \ref{mapas-keplers}, where the top (bottom) panels correspond to Kepler-25 (KE-24), respectively. Left plots were drawn assuming $\alpha=0.5$ while the right graphs are the results obtained considering $\alpha=1.5$. As can be noted, outcomes appear only weakly dependent on the surface density profile, such that we have plotted only the extreme results, and not for $\alpha = 1$, as it is the interpolated plot between $\alpha = 0.5$ and $\alpha = 1.5$. The color code corresponds to the possibility of a pair $(H_0,f)$ to be a solution of equation \eqref{eq16}. Blue colors mean low possibilities while red is associated to higher frequency in the outcomes. 

For Kepler-25, most of the positive results are occur along a broad curve with low values of $H_0$ and large values of the disk flare $f$. This is consistent with the Monte-Carlo simulations presented in \cite{Ramos2017} for the overall near-resonant population and systems with under-determined masses. Notice that the large uncertainty in $m_1$ does not significantly affect the result and the loci of values of $(H_0,f)$ consistent with the observed offset remains fairly restricted.

Results for K2-24 appear less defined which implies that a wide range of disk parameters would lead to the observed resonance offset. In part this is due to the lower, and more easily achieved, value of $\Delta_{2/1}$ but also to particular planetary masses. On one hand, the individual masses are larger than for Kepler-25 which in itself leads to a wider resonance domain. On the other hand, the ratio of $m_2/m_1$ is almost unity which, as seen from equation (\ref{eq15}) also implies larger offset even for low flare values and/or large $H_0$.

\section{Conclusions}

We present an application of the \cite{Ramos2017} model for resonance offset to two planetary systems (Kepler-25 and K2-24) close to a 2/1 MMR with significant observed offset and fairly established planetary masses. We find that the disk parameters necessary to explain the deviation from exact resonance under a laminar type-I migration are similar to those predicted by \cite{Ramos2017}, namely a low disk scale-height and significant flare index. This agreement  indicates that the proposed mechanism could indeed have played a dominant role in determining the observed distribution of (near)-resonant exoplanets.

\ack{This work has been supported by research grants from ANCyT, CONICET and Secyt-UNC. We are 
grateful to IATE and CCAD (UNC) for extensive use of computational facilities.}

\bibliographystyle{iopart-num}
\bibliography{biblio}

\end{document}